\documentclass{ifacconf}

\usepackage{graphicx}      % include this line if your document contains figures
\usepackage{natbib}        % required for bibliography

\usepackage{amsmath,amsfonts}
\usepackage{amssymb}
\usepackage{latexsym}
\usepackage{bm}
\usepackage{mathtools}
\usepackage{empheq}
\usepackage{xcolor}
\usepackage{algorithm}
\usepackage{algpseudocode}
\newcommand{\figref}[1]{Fig.~\ref{#1}}

\newcommand{\asmref}[1]{\textit{{Assumption~\ref{#1}}}}
\newcommand{\propref}[1]{\textit{{Proposition~\ref{#1}}}}

\newcommand{\modifiedI}[1]{#1}
\newcommand{\modifiedII}[1]{#1}
\newcommand{\modifiedIII}[1]{#1}
\newcommand{\modifiedIV}[1]{#1}
\newcommand{\modifiedV}[1]{#1}
\newcommand{\modifiedVI}[1]{#1}
\newcommand{\modifiedVII}[1]{#1}

\begin{document}
\begin{frontmatter}

\title{Language-Aided State Estimation} 

\thanks[footnoteinfo]{
% Sponsor and financial support acknowledgment goes here.
\modifiedI{This work was supported by Grant-in-Aid for Scientific Research (B), No.~25K01254 from JSPS.}
}

\author[Keio]{Yuki Miyoshi} 
\author[Keio]{Masaki Inoue}
\author[Osaka]{Yusuke Fujimoto}

\address[Keio]{Keio University, (e-mail: yukiki0423@keio.jp, minoue@appi.keio.ac.jp)}
\address[Osaka]{The University of Osaka, (e-mail: fujimoto.yusuke.es@osaka-u.ac.jp)}
% \address[Second]{Colorado State University, 
%    Fort Collins, CO 80523 USA (e-mail: author@lamar. colostate.edu)}

\begin{abstract}                % Abstract of 50--100 words
% Natural language data have become increasingly accessible through social networking services and chat platforms.
% These data contain valuable information about human observations, requests, and evaluations, which can support for control and state estimation.
\modifiedIV{
Natural language data, such as text and speech, have become readily available through social networking services and chat platforms.
By leveraging human observations expressed in natural language, this paper addresses the problem of state estimation for physical systems, in which humans act as sensing agents.
}%
% This paper proposes a Language-Aided Particle Filter (LAPF), a particle filter framework that enables the use of natural language observations in state estimation.
% To evaluate its effectiveness, we conducted numerical experiments on river water level estimation, comparing the LAPF with an External DNN-Aided Particle Filter (EDAPF).
% The results demonstrate that the LAPF consistently achieves lower mean squared error than the EDAPF.
\modifiedIV{%
To this end, we propose a Language-Aided Particle Filter (LAPF), a particle filter framework that structures human observations via natural language processing and incorporates them into the update step of the state estimation.
Finally,}
% the proposed LAPF is applied to the state estimation problem at multiple locations along an irrigation canal to demonstrate its effectiveness.
\modifiedV{the LAPF is applied to the water level estimation problem in an irrigation canal and its effectiveness is demonstrated.}
\end{abstract}

\begin{keyword}
% \consult{Five to ten keywords, preferably chosen from the IFAC keyword list.}
\modifiedV{Filtering and smoothing, State estimation, Particle filter, Natural language processing, Human-in-the-loop estimation, Natural language observation}
\end{keyword}

\end{frontmatter}
%===============================================================================

\section{Introduction}

%% [段落1-1]近年、テキストデータが容易に手に入るけど、活用しきれていない
% SNSやチャットサービスの普及により，自然言語形式のデータを容易に取得することが可能になっている．
With increasing use of social networking services and chat platforms, natural language data have become easily available.
% このようなテキストデータには，人間による要望や評価などの情報が多く含まれている．
% These text data contain valuable information such as human requests and evaluations.
\modifiedIV{
These data contain valuable information about human observations, requests, and evaluations, which can be leveraged to improve control accuracy or incorporate human preferences into control systems.}
% しかし，こうした多様な情報を含むテキストデータを解析し，実用的な制御や状態推定に応用する事例は限られている．そのため，入手したテキストデータを効果的に解析し，制御や状態推定に活用するための手法の開発が求められている．
% However, practical applications that exploit text data for control or state estimation remain limited, thus highlighting the need for methods that can effectively utilize such data.
\modifiedIV{
However, practical applications of natural language data in control and state estimation remain limited, underscoring the need for effective methods.}

% %% [段落1-2]自然言語を活用するロボティクス分野の研究
% % 人間からのテキストデータを活用する研究は，ロボティクスの分野において一部行なわれてきた．
% Research on utilizing human-provided text data has been conducted to some extent in the field of robotics.
% % 例えば，文献\cite{matuszek2013learning}では自然言語形式の行動指示をロボットの解釈可能な人工言語形式に変換する手法が提案され，文献\cite{squire2015grounding}では自然言語形式の行動目標からロボットに自律的に行動計画を行なわせる手法が提案されている．
% For example, \cite{matuszek2013learning} proposed a method for translating natural language instructions into a robot-interpretable artificial language, while \cite{squire2015grounding} developed a method that enables robots to autonomously generate action plans from natural language goals.
% % しかし，これらの手法では，自然言語から人工言語への変換をルールベースや小規模な統計モデルにより行なっており，自然言語の多様な表現への対応には限界があった．
% However, these approaches relied on rule-based techniques or small-scale statistical models for the translation from natural language to artificial language, which limited their ability to handle diverse natural language expressions.

%% [段落1-3]LLMの登場により、多様な自然言語も活用できるように
% 近年では，大規模言語モデル(Large Lauguage Models: 以下，LLM)の登場により，複雑なテキストでも文脈情報を捉えて数値化し，柔軟かつ高精度な自然言語解析をすることが可能になっている．
% In recent years, the advent of large language models (LLMs) has brought significant advances on natural language processing.
\modifiedIV{
In recent years, large language models (LLMs) has brought significant advances on natural language processing.}
% They make it possible to capture contextual information from complex text and represent it numerically, thereby enabling flexible and highly accurate analysis.
\modifiedIV{
LLMs not only convert natural language into structured representations but also capture its semantic context, enabling a wide range of applications.}
% このLLMを活用して，自然言語形式の指示文を用いたロボットの制御を行なう研究も現れている．
Several studies leveraging LLMs have emerged that focus on robot control using natural language instructions.
% 例えば，文献\cite{liu2023grounding}では，LLMを用いて自然言語形式の指示文を人工言語形式に変換する手法を提案している． 
For example, \cite{liu2023grounding} proposed a method for translating natural language instructions into an artificial language using LLMs.
% そして，文献\cite{ravichandran2025spine}では，ユーザからの指示文が曖昧で不完全な場合に，LLMを用いてユーザの期待する行動の推論やユーザとの対話をすることで，ロボットがオンラインで行動を決定する手法を提案している．
Furthermore, \cite{ravichandran2025spine} introduced a framework for handling ambiguous or incomplete user instructions.
This framework employs LLMs to infer the intended actions or to engage in dialogue with the user, thereby allowing the robot to make decisions online.
% そのほかにも，文献\cite{miyaoka2024chatmpc, wu2025instructmpc}では，LLMをモデル予測制御に統合し，自然言語形式の指示や状況説明を用いて制御パラメータの調整を行なう手法を提案している．
In addition, \cite{miyaoka2024chatmpc} and \cite{wu2025instructmpc} developed approaches that integrate LLMs into model predictive control,  thus enabling the adjustment of control parameters based on natural language instructions and situational descriptions.

%% [段落1-4]本研究では、LLMを状態推定に対して適用することを考える
% 本稿では，テキストデータを状態推定に活用する手法について検討する．
% In this paper, we investigate methods for utilizing text data in the field of state estimation.
\modifiedIV{
In this paper, we leverage natural language data within state estimation for physical systems, in what is, to the best of the authors' knowledge, the first attempt to explore this integration.}
% 具体的には，自然言語形式の観測情報を返す人間をセンサと見なし，動的システムの内部状態の推定問題に取り組む．
% Specifically, we address the problem of estimating the internal states of a dynamical system by treating humans providing natural language observations as sensors.
\modifiedIV{
To this end, we first design a language model that processes human observations and interprets them as probability distributions over a part of the state.
Then, treating humans as sensing agents, we develop a state estimation method in which the natural language observations are incorporated into the update step of the particle filter.}
% そのアプローチとして，人間による観測操作をモデル化し，そのモデルに基づいた状態推定手法を新たに考案する．
% In our approach, we model the human observation process and develop a state estimation method based on the model.

%% [段落1-5]状態推定の代表的な研究
% 状態推定の代表的な手法には，カルマンフィルタ\cite{kalman1960new}やパーティクルフィルタ\cite{gordon1993novel}などがあり，対象の状態を確率的に推定するために広く用いられている．
% Representative methods for state estimation include the Kalman filter proposed by \cite{kalman1960new} and the particle filter introduced by \cite{gordon1993novel}, which are widely used to probabilistically estimate the states of target systems.
% 近年では，深層学習を組み合わせた状態推定手法も研究されている．
% In recent years, state estimation methods that incorporate deep learning have also been developed.
\modifiedIV{
Several related works have applied Artificial Intelligence (AI)–based techniques, including but not limited to natural language processing, to the problem of state estimation.}
% 例えば，文献\cite{revach2022kalmannet}では，カルマンゲインの計算をNeural Networkを使って行なうKalmanNetが提案され，モデルが不確かな場合や非線形の場合において従来のカルマンフィルタを上回る性能を発揮することを実証した．
% For example, \cite{revach2022kalmannet} proposed KalmanNet, which computes the Kalman gain using a neural network, and demonstrated that this approach can outperform classical Kalman filtering methods under model uncertainty and nonlinear dynamics.
\modifiedIV{
For example, \cite{revach2022kalmannet} incorporated deep neural networks into a Kalman filter variant, where the Kalman gain is updated based on the network output.}
% 文献\cite{ghosh2024danse}では，観測データを用いて，対象システムのモデリングと事後分布の計算をRNNにより行ない，教師なし学習を用いた状態推定方法DANSEを提案した．この手法では，対象システムのモデルが不明な場合でも非線形な状態推定を可能にし，RNNの学習に真の状態データを必要としない．
% In addition, \cite{ghosh2024danse} proposed DANSE, an unsupervised state estimator that uses recurrent neural networks to model system dynamics and compute posterior state distributions directly from observations.
% They showed that this approach enables nonlinear state estimation when no explicit system model is available, without requiring ground-truth state trajectories for training.
\modifiedIV{
In addition, \cite{ghosh2024danse} proposed an unsupervised state estimation approach that uses a recurrent neural network to model system dynamics and infer posterior state distributions directly from observations.}

\modifiedV{
\textbf{Theoretical Contribution:}
This paper introduces the Language-Aided Particle Filter (LAPF), a state estimation framework that incorporates human natural language observations via natural language processing techniques.}

\modifiedV{
\textbf{Technological Contribution:}
It demonstrates the practical robustness of the LAPF against out-of-domain human observation data that were not used during training.}

%% [段落1-7]本稿の構成
% 本稿の構成を述べる．
The remainder of this paper is organized as follows.
% \ref{sec:Problem Setting}節で本稿における問題設定,
Section~\ref{sec:Problem Setting} describes the problem setting 
% considered in this study. 
\modifiedI{of the state estimation incorporating natural language observations.}
% \ref{sec:Language-Aided Particle Filter}節で提案手法の LAPF,
Section~\ref{sec:Language-Aided Particle Filter} presents
% the proposed method, the LAPF.
\modifiedI{the LAPF as the solution to the state estimation problem.}
% \ref{sec:Numerical Experiment}節で数値実験,
Section~\ref{sec:Numerical Experiment} 
% reports numerical experiments,
\modifiedIV{presents numerical experiments to demonstrate the LAPF,}
% \ref{sec:Conclusion}節で結論を述べる．
and Section~\ref{sec:Conclusion} concludes this paper.

%% [段落1-8]本稿の表記
% 本稿における表記について述べる．
% The notations used in this paper are defined as follows.
\modifiedI{Notation:}
% 実数の集合は$\mathbb{R}$，テキストの集合は$\mathbb{T}$，空集合は$\phi$と表す．
% The set of real numbers is denoted by $\mathbb{R}$, the set of texts by $\mathbb{T}$, and the empty set by $\phi$.
\modifiedI{The symbol $\mathbb{R}$ denotes the set of real numbers, the symbol $\mathbb{T}$ denotes the set of natural language texts, and the symbol $\phi$ denotes the empty set.}
% テキストの集合$\mathbb{T}$は，「今年の連合講演会は名古屋で開催される。」のようなテキストを要素にもつ集合である．
% The set of texts $\mathbb{T}$ consists of elements elements such as
\modifiedI{Elements of the set $\mathbb{T}$ include, for example,
``IFAC2026 will be held in Busan.''}
% $a \times a$ 次元 の単位行列は$I_a$，$a$行の零ベクトルは$0_a$と表す．
% The $a \times a$ identity matrix is denoted by $I_a$, and the zero vector of length $a$ by $0_a$.
\modifiedI{The symbol $I_n$ denote the $n \times n$ identity matrix, and the symbol $0_n$ denotes the $n$-dimensional zero vector.}
% 平均$\mu$，共分散行列$\Sigma$の正規分布は$\mathcal{N}(\mu, \Sigma)$と表す．
% A normal distribution with mean $\mu$ and covariance matrix $\Sigma$ is denoted by $\mathcal{N}(\mu, \Sigma)$.
\modifiedI{The symbol $\mathcal{N}(\mu, \Sigma)$ denotes the normal distribution with mean $\mu$ and covariance matrix $\Sigma$.}
% 確率分布$p(x)$からサンプリングした値$x$は，$x \sim p(x)$と表す．
% A value $x$ sampled from a probability distribution $p(x)$ is denoted by $x \sim p(x)$.
\modifiedI{The expression $x \sim p(x)$ denotes sampling $x$ from the probability distribution $p(x)$.}
% 時刻$T$までのデータ列$\{x_i\}_{i=1}^T$は$x_{1:T}$と表す．
% A sequence of data up to time $T$, $\{x_i\}_{i=1}^T$, is denoted by $x_{1:T}$.
\modifiedI{The symbol $x_{1:T}$ denotes the time sequence $\{x_1, x_2, \dots, x_T\}$.}
% ディラックのデルタ関数は$\delta(\cdot)$と表す．
% Finally, the Dirac delta function is denoted by $\delta(\cdot)$.
\modifiedI{Finally, the symbol $\delta(\cdot)$ denotes the Dirac delta function.}

\section{Problem Setting} \label{sec:Problem Setting}

%% [段落2-1]本節の概要
% 本稿では，電気的・機械的な素子などにより測定する一般的なセンサに対して，人間を自然言語形式の観測情報を回答するセンサとして用いることを考える．
% In this paper, in contrast to conventional sensors that measure physical quantities based on electrical or mechanical components, we consider humans as sensors providing natural language observations.
\modifiedI{In this paper, we address the state estimation problem for dynamical systems using human observations expressed in natural language.}
% 本稿ではこれを人センサと呼称する．
% We refer to this type of sensor as a human sensor.
\modifiedI{In particular, to clearly distinguish such observations from conventional \textit{physical} sensors that directly measure physical quantities, we refer to them collectively as a \textit{human} sensor.}
% 本節では，まず本稿で想定する対象システムのモデルについて述べ，最後に人センサのモデルについて議論する．
% In this section, we first describe the model of the target system considered in this study, and then discuss the model of the human sensor.
\modifiedI{In this section, we first describe the model of the plant system, and then discuss the model of the human sensor.}

\subsection{\modifiedI{Plant} System $\mathcal{P}$}

%% [段落2-1-1]対象システム
% 対象システムとして，次式の離散時間の状態方程式を考える．
% The target system is modeled by the following discrete-time state equation:
\modifiedI{The dynamics of the plant system, which is the subject of state estimation, are described by the \modifiedVI{following} discrete-time state equation:}
\begin{align}
    \mathcal{P}: \; x_k = f(x_{k-1}, w_{k}), \quad \forall \; k \ge 1\modifiedI{,} \label{eq:StateEquation 1}
\end{align}
% ここで，$x_k \in \mathbb{R}^n$は時刻$k$における状態，$w_k \in \mathbb{R}^{\modifiedIV{\ell}}$は時刻$k$におけるプロセス雑音である．関数 $f: \mathbb{R}^n \times \mathbb{R}^{\modifiedIV{\ell}} \to \mathbb{R}^n$ は，状態$x_{k-1}$とプロセス雑音$w_k$から次時刻$k$の状態$x_k$を出力する状態遷移関数である．
% Here, $x_k \in \mathbb{R}^n$ denotes the state at time $k$, and $w_k \in \mathbb{R}^{\modifiedIV{\ell}}$ denotes the process noise at time $k$.
\modifiedI{where $k$ denotes the discrete-time, $x_k \in \mathbb{R}^n$ denotes the state, $w_k \in \mathbb{R}^{\modifiedIV{\ell}}$ denotes the process noise
\modifiedIV{drawn from a distribution $\mathcal{W}$, and}}
% The function $f: \mathbb{R}^n \times \mathbb{R}^{\modifiedIV{\ell}} \to \mathbb{R}^n$ denotes the state transition function that maps the previous state $x_{k-1}$ and the process noise $w_k$ to the current state $x_k$.
\modifiedI{
% The function
$f: \mathbb{R}^n \times \mathbb{R}^{\modifiedIV{\ell}} \to \mathbb{R}^n$ denotes a nonlinear function.}

\subsection{human sensor $\mathcal{S}_{\mathrm{H}}$} \label{subsec:HumanSensor}

\begin{figure}[t]
		\centering
		\includegraphics[width=0.85\linewidth]{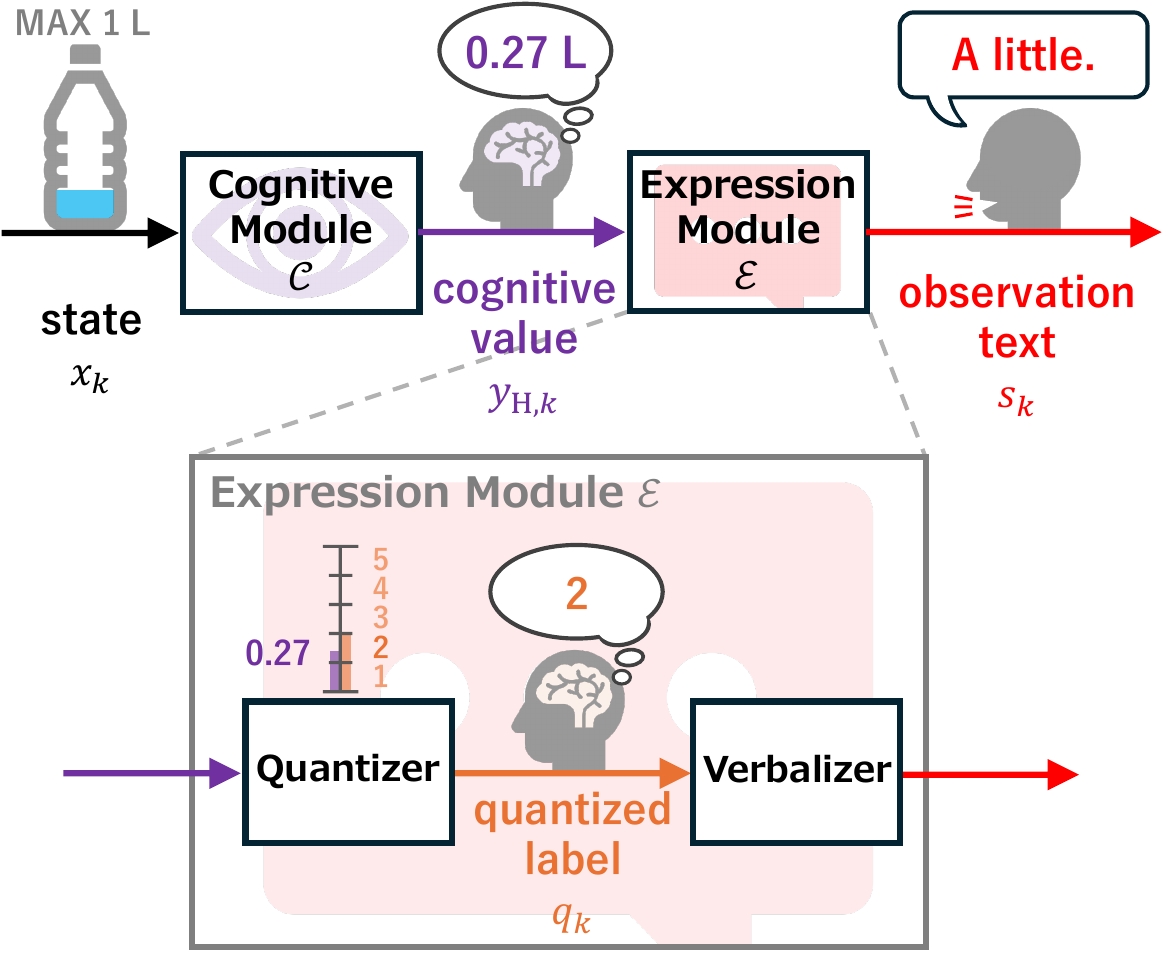}
		\caption{
			Human Sensor Observation Process (for $m=5$)
			\label{fig:Human Sensor Observation Process}
			}
\end{figure}

%% [段落2-2-1]Cognitive ModuleとExpression Moduleのカスケード構造
% 人センサのモデル化を行なう上で，まず，状態$x_k$を観測した人間が自然言語形式の観測情報$s_k$を報告する際，その脳内でどのような処理が行なわれているかを考える．
To model the human sensor, 
% we begin by considering the cognitive process that occurs when an observer of the state $x_k$ reports a natural language observation $s_k$.
\modifiedI{we begin by focusing on the observation process through which a human sensing agent perceives a part of the state $x_k$ and reports it as a natural language observation $s_k$.}
% ここで，本小節では説明のため，観測者は一人とし，その観測者が報告するテキスト$s_k$も一つ，つまり$s_k \in \mathbb{T}$という単純な状況で考える．
For simplicity, in this subsection we restrict our attention to the case of a single
% observer
\modifiedI{agent}
who produces a single text $s_k \in \mathbb{T}$.
% \figref{fig:Human Sensor Observation Process}に示すように，人センサにおける観測プロセスは，\textbf{Cognitive Module} $\mathcal{C}$ と\textbf{Expression Module} $\mathcal{E}$ のカスケード構造でモデル化する．
As illustrated in \modifiedI{the upper block of} \figref{fig:Human Sensor Observation Process},
% in this paper
we model the observation process of the human sensor as a cascade of two modules: the
% Cognitive Module
\modifiedI{cognitive module} $\mathcal{C}$ and the
% Expression Module
\modifiedI{expression module} $\mathcal{E}$.
% \begin{figure}[t]
% 		\centering
% 		\includegraphics[width=1\linewidth]{figure/Image_of_HumanSensor_by_Text_Eng.pdf}
% 		\caption{
% 			Human Sensor Observation Process
% 			\label{fig:Human Sensor Observation Process}
% 			}
% \end{figure}
% 具体的には，観測プロセスを，(i)状態$x_k$を知覚し，脳内で認知値$y_{\mathrm{H}, k} \in \mathbb{R}$を取得，(ii)認知値$y_{\mathrm{H}, k}$を観測テキスト$s_k$として表現，の2つのステップでモデル化する．
% Specifically, we model this process as consisting of two stages: (i) the
% % observer
% \modifiedI{agent}
% perceives the state $x_k$ and forms an internal cognitive value $y_{\mathrm{H},k} \in \mathbb{R}$, and (ii) this cognitive value is expressed as the observation text $s_k$.
\modifiedIV{
In $\mathcal{C}$, the agent perceives the state $x_k$ and forms an internal cognitive value $y_{\mathrm{H},k} \in \mathbb{R}$.
Then, in $\mathcal{E}$, this cognitive value is expressed as the observation text $s_k$.}
% 以下では，Cognitive Module $\mathcal{C}$とExpression Module $\mathcal{E}$のモデルについて述べる．
In the following, we describe the models of
% the \modifiedI{cognitive module} $\mathcal{C}$ and the \modifiedI{expression module} $\mathcal{E}$.
\modifiedIV{$\mathcal{C}$ and $\mathcal{E}$.}

%% [段落2-2-2]Cognitive Module
% Cognitive Module $\mathcal{C}$は，次式の観測方程式で表現できると考える．
% The Cognitive Module $\mathcal{C}$ is modeled by the following observation equation:
\modifiedI{In the cognitive module $\mathcal{C}$, the internal cognitive value $y_{\mathrm{H}, k}$ is determined by the state $x_k$ as}
\begin{align}
    \mathcal{C}: \; y_{\mathrm{H}, k} = h_{\mathrm{H}} ( x_k, v_{\mathrm{H}, k} ), \quad \forall \; k \ge 1 \modifiedI{,} \label{eq:Recognizer of HumanSensor 1}
\end{align}
% ここで，$v_{\mathrm{H}, k} \in \mathbb{R}^{r'}$は観測者の認知上の観測雑音である．関数 $h_{\mathrm{H}}: \mathbb{R}^n \times \mathbb{R}^{r'} \to \mathbb{R}$ は，状態$x_k$と認知雑音$v_{\mathrm{H}, k}$から認知値$y_{\mathrm{H}, k}$を出力する観測関数である．
% Here,
\modifiedI{where} $v_{\mathrm{H}, k} \in \mathbb{R}^{\modifiedIV{r}}$ denotes the cognitive measurement noise
\modifiedIV{drawn from a distribution $\mathcal{V_{\mathrm{H}}}$, and}
% of the observer.
% The function $h_{\mathrm{H}}: \mathbb{R}^n \times \mathbb{R}^{\modifiedIV{r}} \to \mathbb{R}$ denotes the observation function that maps the state $x_k$ and the cognitive noise $v_{\mathrm{H}, k}$ to the cognitive value $y_{\mathrm{H}, k}$.
\modifiedI{% The function
$h_{\mathrm{H}}: \mathbb{R}^n \times \mathbb{R}^{\modifiedIV{r}} \to \mathbb{R}$ denotes a nonlinear function.}

%% [段落2-2-3]Expression Module
% \figref{fig:Language Generator Process}に示すように，Expression Module$\mathcal{E}$はQuantizerとVerbalizerで構成されるとする．
As illustrated in 
% \figref{fig:Language Generator Process},
\modifiedI{the lower block of \figref{fig:Human Sensor Observation Process},}
% the \modifiedI{expression module} $\mathcal{E}$ is modeled as consisting of two components: a Quantizer and a Verbalizer.
\modifiedI{we further model the expression module $\mathcal{E}$ as being composed of a quantizer and a verbalizer.}
% \begin{figure}
% 		\centering
% 		\includegraphics[width=1\linewidth]{figure/Image_of_ExpressionModule_Eng.pdf}
% 		\caption{
% 			Expression Module Process (for $m=5$)
% 			\label{fig:Language Generator Process}
% 		}
% \end{figure}
% Figの構造
\modifiedI{In other words, the model
% in the figure
\modifiedIV{$\mathcal{E}$} assumes that a quantization process mediates the verbalization of cognitive information by human agents.}
% このように考えられる理由は，観測者による言語表現能力には限界があり，認知値$y_{\mathrm{H}, k}$に対して細かく言語化することはできないためである．
% This modeling is justified by the limitations of human linguistic ability, which make it impossible for an \modifiedI{agent} to verbalize the cognitive value $y_{\mathrm{H},k}$ with sufficiently fine granularity.
\modifiedI{This reflects the limitation of human linguistic ability, which make it impossible for an agent to verbalize the cognitive value $y_{\mathrm{H},k}$ at high resolution.}
% 例えば，\figref{fig:Language Generator Process}において，認知値$y_{\mathrm{H}, k}$が$0.27~\mathrm{L}$の場合と$0.28~\mathrm{L}$の場合に対して，観測者がそれらを区別し，それぞれ異なる観測テキスト$s_k$を回答することは困難である．
For example, as shown in
% \figref{fig:Language Generator Process}, it is unrealistic to expect an \modifiedI{agent} to produce distinct observation texts $s_k$ for cognitive values of $0.27~\mathrm{L}$ and $0.28~\mathrm{L}$.
\modifiedI{\figref{fig:Human Sensor Observation Process}, it is reasonable to assume that human agents cannot distinguish between the cognitive values $0.27~\mathrm{L}$ and $0.28~\mathrm{L}$ when generating observation texts.}
% そのため，観測者が認知値$y_{\mathrm{H}, k}$を表現する際，$y_{\mathrm{H}, k}$から直接言語化を行なうのではなく，$y_{\mathrm{H}, k}$を$m$段階の値に量子化した量子ラベル$q_k \in \{1, \dots, m\}$から言語化を行なっていると考えることは妥当である．
% Therefore, it is reasonable to assume that the \modifiedI{agent} verbalizes not the cognitive value $y_{\mathrm{H},k}$ but instead a quantized label $q_k \in \{1, \dots, m\}$, which is derived by quantizing $y_{\mathrm{H},k}$ into $m$ discrete levels.
\modifiedI{We further assume that the quantizer has $m$ quantization levels and outputs a corresponding quantization label $q_k \in \{1, \dots, m\}$.}
% したがって，Expression Module$\mathcal{E}$は，次式で記述できる．
% From this modeling,
\modifiedI{In summary,} the \modifiedI{expression module} $\mathcal{E}$
% can be formulated
\modifiedI{is described} as follows:
\begin{subequations} \label{eq:LanguageGenerater of HumanSensor 1}
    \begin{empheq}[left={\mathcal{E}: \empheqlbrace}]{alignat=2}
        &q_k = Q_m (y_{\mathrm{H}, k}),& \quad &\forall \; k \ge 1, \label{subeq:Quantizer of LanguageGenerater 1} \\
        &s_k \sim \mathrm{Label 2 Prob}(q_k),& \quad &\forall \; k \ge 1 \modifiedI{,} \label{subeq:LanguageGenerater of LanguageGenerater 1}
    \end{empheq}
\end{subequations}
% ここで，関数 $Q_m: \mathbb{R} \to \{1, \dots , m\}$は，認知値$y_{\mathrm{H}, k}$を$m$段階の量子ラベル$q_k$に量子化する関数である．
% Here, 
\modifiedI{where} the function $Q_m: \mathbb{R} \to \{1, \dots , m\}$ denotes the quantizer that maps the cognitive value $y_{\mathrm{H}, k}$ to a quantized label $q_k$ with $m$ discrete levels.
% また，関数$\mathrm{Label 2 Prob}$は，量子ラベル$q_k$に対応した観測テキスト$s_k$の確率分布を生成する関数である．
The function $\mathrm{Label2Prob}$ generates a probability distribution over observation texts $s_k$ corresponding to the quantized label $q_k$.
% ただし，この関数$\mathrm{Label 2 Prob}$は，人間の言語化を数式で記述するために導入した抽象的な関数である．
Note that $\mathrm{Label2Prob}$ is introduced as an abstract function to formally describe the process of human
% linguistic expression.
\modifiedI{observation.}
\modifiedI{While we assume the existence of $\mathrm{Label2Prob}$, its explicit functional form is not used in the subsequent discussion.}
% \eqref{subeq:LanguageGenerater of LanguageGenerater 1}式では，観測テキスト$s_k$の生成を，関数$\mathrm{Label 2 Prob}$で得られる確率分布からサンプリングすることで表現している．
Equation~\eqref{subeq:LanguageGenerater of LanguageGenerater 1} describes the generation of the observation text $s_k$ as a sample drawn from the distribution given by $\mathrm{Label2Prob}$.

%% [段落2-2-4]量子化関数の定義
% 最後に，量子化関数$Q_m$の定義を述べる．
Finally, we
% define
\modifiedI{state the details of} the quantization function $Q_m$.
% 認知値$y_{\mathrm{H}, k}$の取り得る範囲を$\Lambda = [y_{\mathrm{H}, \mathrm{min}}, y_{\mathrm{H}, \mathrm{max}}]$とする.
% Let $\Lambda = [y_{\mathrm{H}, \mathrm{min}}, y_{\mathrm{H}, \mathrm{max}}]$ denote the range of possible cognitive values $y_{\mathrm{H},k}$.
\modifiedI{Let $\Lambda \subset \mathbb{R}$ denote the value range of $y_{\mathrm{H},k}$, and further let $\Lambda = [y_{\mathrm{H}, \mathrm{min}}, y_{\mathrm{H}, \mathrm{max}}]$.}
% また，$\Lambda$上の$m$個の量子化区間$\{\Lambda_i\}_{i=1}^m$は，次式を満たすとする．
We partition $\Lambda$ into $m$ quantization intervals $\{\Lambda_i\}_{i=1}^m$ that satisfy the following conditions:
\begin{subequations} \label{eq:QuantizerRange}
    \begin{gather}
        \cup_{i=1}^m \Lambda_i =\Lambda, \label{subeq:Const of QuantizerRange 1} \\
        \Lambda_i \cap \Lambda_j = \phi, \quad \forall i \ne j. \label{subeq:Const of QuantizerRange 2}
    \end{gather}
\end{subequations}
% このとき，量子化関数$Q_m$を次式で定義する．
The quantization function $Q_m$ is then defined as
\begin{align}
		Q_m(y_{\mathrm{H}, k}) \coloneq i, \quad \mathrm{if} \; y_{\mathrm{H}, k} \in \Lambda_i. \label{eq:QuantizerFunc}
\end{align}

% \modifiedVI{
% Within this problem setting, we address the following state estimation problem:
% \begin{prob}
%     Given a sequence of observation texts $\{s_k\}_{k=1}^N$, and under a predefined set of quantization intervals $\{\Lambda_i\}_{i=1}^m$, estimate the posterior distribution of the state $x_k$ for each time step $k$.
% \end{prob}
% }
\modifiedVII{
Within this problem setting, we address the following state estimation problem:
\begin{prob}
    Consider the plant system $\mathcal{P}$, given in \eqref{eq:StateEquation 1}, and the human sensor $S_{\mathrm{H}} = \{\mathcal{C}, \mathcal{E}\}$, given in \eqref{eq:Recognizer of HumanSensor 1}, \eqref{eq:LanguageGenerater of HumanSensor 1}, \eqref{eq:QuantizerRange}, and \eqref{eq:QuantizerFunc}.
    Given the distribution of the initial state $\pi(x_0)$ and the sequence of observation texts $s_{1:k}$,
    % and under a predefined set of quantization intervals $\{\Lambda_i\}_{i=1}^m$,
    estimate the distribution of the state $\pi (x_k)$.
\end{prob}
}

\section{Language-Aided Particle Filter} \label{sec:Language-Aided Particle Filter}

%% [段落3-1]概要
% 本節では，\ref{sec:Problem Setting}節で述べたモデルのもとで，パーティクルフィルタの理論に基づいて状態推定を行なう手法である LAPF について述べる．
In this section, we describe the LAPF, \modifiedIV{the particle filter framework of}
% a state estimation method based on the particle filter framework using the modeling assumptions presented in Section~\ref{sec:Problem Setting}.
\modifiedI{using natural language observations $s_k$.}

%% [段落3-2]パーティクルフィルタの概要
% パーティクルフィルタとは，非線形・非ガウスの状態空間モデルに対しても適用可能な状態推定手法である．
% The particle filter is a state estimation method that
\modifiedI{We briefly review the particle filter, as proposed in \cite{gordon1993novel}, which} can be applied to nonlinear and non-Gaussian state-space models.
% パーティクルフィルタでは，事後分布 $\pi(x_k | s_{1:k})$ として，近似事後分布$\tilde{\pi}(x_k | s_{1:k})$ を $N_{\mathrm{p}}$ 個のパーティクル $\{x_{k | k-1}^i\}_{i=1}^{N_{\mathrm{p}}}$ とその重み $\{\alpha_k^i\}_{i=1}^{N_{\mathrm{p}}}$ により計算する．
% It
\modifiedI{The particle filter} approximates the posterior distribution $\pi(x_k | s_{1:k})$ with an empirical distribution $\tilde{\pi}(x_k | s_{1:k})$, represented by $N_{\mathrm{p}}$ particles $\{x_{k | k-1}^i\}_{i=1}^{N_{\mathrm{p}}}$ and their corresponding weights $\{\alpha_k^i\}_{i=1}^{N_{\mathrm{p}}}$,
\modifiedIV{where $N_{\mathrm{p}}$ is the number of particles.}

%% [段落3-3]具体的な手順
% 具体的な手順を簡単に述べる．
The procedure of the particle filter
% can be
\modifiedI{is} summarized as follows.
% まず，前時刻$k-1$の近似事後分布$\tilde{\pi}(x_{k-1} | s_{1:k-1})$からサンプリングを行ない，$N_{\mathrm{p}}$個のパーティクル$\{x_{k-1 | k-1}^i\}_{i=1}^{N_{\mathrm{p}}}$を取得する．
% First,
\modifiedII{At the first step,} $N_{\mathrm{p}}$ particles $\{x_{k-1 | k-1}^i\}_{i=1}^{N_{\mathrm{p}}}$ are drawn from the empirical distribution $\tilde{\pi}(x_{k-1} | s_{1:k-1})$.
% つぎに，対象システム$\mathcal{P}$の状態方程式\eqref{eq:StateEquation 1}を用いて，パーティクル$\{x_{k-1 | k-1}^i\}_{i=1}^{N_{\mathrm{p}}}$を次式で時間更新する．
Next, these particles are propagated forward according to the state equation
% of the system $\mathcal{P}$:
\modifiedI{given in \eqref{eq:StateEquation 1} as}
\begin{align}
    x_{k | k-1}^i = f(x_{k-1 | k-1}^i, w_k^i), \quad \forall i \in \{1, \dots, N_{\mathrm{p}}\}, \label{eq:PredictionStep of ParticleFilter1}
\end{align}
% ここで$w_k^i$ は，システム雑音 $w_k$ の確率分布$p(w_k)$に従って，各パーティクル$x_{k-1 | k-1}^i$毎に独立にサンプリングされた乱数である．
where
% $w_k^i$ denotes a sample of the process noise $w_k$ drawn independently for each particle from its distribution $p(w_k)$.
\modifiedII{$w_k^i$ denote i.i.d. samples drawn from the process noise distribution
% $p(w_k)$
\modifiedIV{$\mathcal{W}$}.}
% そして，観測情報$s_k$が得られた後，各パーティクル$x_{k | k-1}^i$に対して，次式で重み$\alpha_k^i$を計算する．
% After the observation $s_k$ is obtained, 
\modifiedII{At the second step,}
\modifiedI{given an observation of the state, such as \modifiedIV{observation text} $s_k$ in the problem addressed in this paper,}
% the weights
\modifiedVI{each weight} $\alpha_k^i$ of the predicted particles \modifiedVI{$\{x_{k | k-1}^i\}_{i=1}^{N_{\mathrm{p}}}$ is} 
% are
updated
% based on the likelihood function:
\modifiedVI{according to the likelihood $p(s_k \mid x_{k\mid k-1}^i)$ as}
\begin{align}
    \alpha_k^i = \frac{p (s_k | x_{k | k-1}^i)}{\sum_{j=1}^{N_{\mathrm{p}}} p (s_k | x_{k | k-1}^j)}, \quad \forall i \in \{1, \dots, N_{\mathrm{p}}\}, \label{eq:UpdateStep of ParticleFilter1}
\end{align}
% ここで，$p (s_k | x_{k | k-1}^i)$は尤度であり，状態が $x_{k | k-1}^i$ である場合に観測情報 $s_k$が得られる確率である．
where $p(s_k | x_{k | k-1}^i)$ denotes the likelihood of $s_k$ given the state $x_{k | k-1}^i$.
% 最後に，パーティクル $\{x_{k | k-1}^i\}_{i=1}^{N_{\mathrm{p}}}$ とその重み $\{\alpha_k^i\}_{i=1}^{N_{\mathrm{p}}}$を用いて，近似事後分布 $\tilde{\pi}(x_k | s_{1:k})$ を次式で計算する．
% Finally,
\modifiedII{At the last step,}
the empirical distribution $\tilde{\pi}(x_k | s_{1:k})$ is represented as a weighted sum of delta functions:
\begin{align}
    \tilde{\pi} (x_k | s_{1:k}) = \sum_{i=1}^{N_{\mathrm{p}}} \alpha_k^i \; \delta(x_k - x_{k | k-1}^i). \label{eq:UpdateStep of ParticleFilter2}
\end{align}
% パーティクルフィルタは，これらの処理を各時刻$k$で逐次的に繰り返すことで状態推定を行なう手法である．
By 
% repeating these steps at each time step $k$,
\modifiedII{iterating these steps to obtain the empirical distribution $\tilde{\pi}(x_k | s_{1:k})$ at each time $k$,} the particle filter provides an approximation of the posterior distribution \modifiedI{$\pi(x_k | s_{1:k})$}.

%% [段落3-4]重要なのは尤度
% \ref{sec:Problem Setting}節で述べたモデルに対してパーティクルフィルタを適用する上で肝要となるのは，\eqref{eq:UpdateStep of ParticleFilter1}式に現れる尤度$p (s_k | x_{k | k-1}^i)$の計算である．
The key step in applying the particle filter to the state-space model
% in Section~\ref{sec:Problem Setting}
\modifiedIV{defined by \eqref{eq:StateEquation 1}, \eqref{eq:Recognizer of HumanSensor 1}, and \eqref{eq:LanguageGenerater of HumanSensor 1}} is the computation of the likelihood $p(s_k | x_{k | k-1}^i)$ that appears in \eqref{eq:UpdateStep of ParticleFilter1}.
% 本節では，まず\ref{subsec:HumanSensor}節で述べたような単一の人センサ$\mathcal{S}_{\mathrm{H}}$を使用する場合の尤度$p (s_k | x_{k | k-1}^i)$の計算方法について述べる．
% In this section,
\modifiedIV{In the following two subsections,} we first
% describe how to compute
\modifiedII{provide the computation of} this likelihood in the case of a single human sensor $\mathcal{S}_{\mathrm{H}}$, as considered in Subsection~\ref{subsec:HumanSensor}.
% そして，その拡張として，複数の人センサ$\mathcal{S}_{\mathrm{H}}$を使用する場合の尤度の計算方法についても述べる．
We then extend this approach to the case of multiple human sensors $\mathcal{S}_{\mathrm{H}}$.

\subsection{The case of a single human sensor} \label{subsec:How to get Likelihood}

%% [段落3-1-1]概要
% 本小節では，単一の人センサ$\mathcal{S}_{\mathrm{H}}$を使用する場合の尤度$p (s_k | x_{k | k-1}^i)$の計算方法について述べる．
% In this subsection, we
% % describe how to compute
% \modifiedII{provide the computation of} the likelihood $p(s_k | x_{k | k-1}^i)$ in the case of a single human sensor $\mathcal{S}_{\mathrm{H}}$.

%% 新[段落3-1-2]p(s|x)の計算
\modifiedII{
% まず，p(s|x)をq_kを使って記述する
First, we 
% represent
\modifiedIV{rewrite} the likelihood $p(s_k | x_{k | k-1}^i)$
% in terms of
\modifiedIV{using} the quantized label $q_k$. 
% その準備のために，量子ラベルの確率分布に対して次の仮定をおく
For this purpose, we assume the following prior distribution for the quantized label $q_k$:
\begin{assum} \label{asm:PriorDistributuon of q}
		% 量子ラベル$q_k$の確率分布$p(q_k)$は次式で記述できると仮定する．
		The quantized label $q_k$ is assumed to follow a uniform distribution, i.e.,
    \begin{align}
        p(q_k^j) = \frac{1}{m}, \quad \forall j \in \{1, \dots, m\}. \label{eq:Asm of PriorDistributuon of q}
    \end{align}
\end{assum}
Under this assumption, the following proposition holds:
\begin{prop} \label{prop:likelihood-factorization}
    Under \asmref{asm:PriorDistributuon of q},
    % the likelihood $p (s_k | x_{k | k-1}^i)$ satisfies
    \modifiedIV{it holds that}
    \begin{align}
        p(s_k | x_{k | k-1}^i)
        \propto \sum_{j=1}^m p(q_k^j | s_k)\, p(q_k^j | x_{k | k-1}^i).
        \label{eq:How to get Likelihood3}
    \end{align}
\end{prop}
\begin{pf}
    Note that the likelihood $p(s_k | x_{k | k-1}^i)$ is obtained by marginalizing the joint distribution $p(s_k, q_k | x_{k | k-1}^i)$ with respect to $q_k$.
    In addition, recall the definition of conditional probability and the fact that the observation $s_k$ depends only on $q_k$ as follows from \eqref{subeq:LanguageGenerater of LanguageGenerater 1}.
    Then, it follows that
    \begin{align}
        p (s_k | x_{k | k-1}^i) = \sum_{j = 1}^m \; p (s_k | q_k^j) \; p (q_k^j | x_{k | k-1}^i). \label{eq:How to get Likelihood1}
    \end{align}
    By applying Bayes' theorem to $p(s_k | q_k^j)$ in \eqref{eq:How to get Likelihood1}, and noting that $p(s_k)$ is independent of the particle index $i$ and cancels out in the computation of the weights $\alpha_k^i$ in \eqref{eq:UpdateStep of ParticleFilter1}, it follows that
    \begin{align}
        p (s_k | x_{k | k-1}^i) &= \sum_{j = 1}^m \; \frac{p (q_k^j | s_k) \; p (s_k) }{p(q_k^j)}  \; p (q_k^j | x_{k | k-1}^i) \nonumber \\
        &\propto \sum_{j = 1}^m \; \frac{p (q_k^j | s_k)}{p(q_k^j)}  \; p (q_k^j | x_{k | k-1}^i). \label{eq:How to get Likelihood2}
    \end{align}
    By substituting \eqref{eq:Asm of PriorDistributuon of q} into \eqref{eq:How to get Likelihood2}, it follows that
    \begin{align}
        p (s_k | x_{k | k-1}^i) \propto m \sum_{j = 1}^m \; p (q_k^j | s_k)  \; p (q_k^j | x_{k | k-1}^i) \label{eq:How to get Likelihood4}
    		% &\propto \sum_{j = 1}^m \; p (q_k^j | s_k)  \; p (q_k^j | x_{k | k-1}^i). \label{eq:How to get Likelihood4}
    \end{align}
    This completes the proof of
    % \thmref{prop:likelihood-factorization}.
    \modifiedVI{\propref{prop:likelihood-factorization}.}
\end{pf}
% したがって，$p (q_k^j | s_k)$ と $p (q_k^j | x_{k | k-1}^i)$ を計算することで，尤度$p (s_k | x_{k | k-1}^i)$を計算することができる．
\modifiedIV{The \propref{prop:likelihood-factorization} states that}
% Therefore, 
the likelihood $p(s_k | x_{k | k-1}^i)$ is obtained by computing $p(q_k^j | s_k)$ and $p(q_k^j | x_{k | k-1}^i)$.
% 以下では，$p (q_k^j | s_k)$ と $p (q_k^j | x_{k | k-1}^i)$ の計算方法について述べる．
In the following, we provide the computation of \modifiedIV{them.}
% each term.
}

\begin{figure}
		\centering
		\includegraphics[width=1\linewidth]{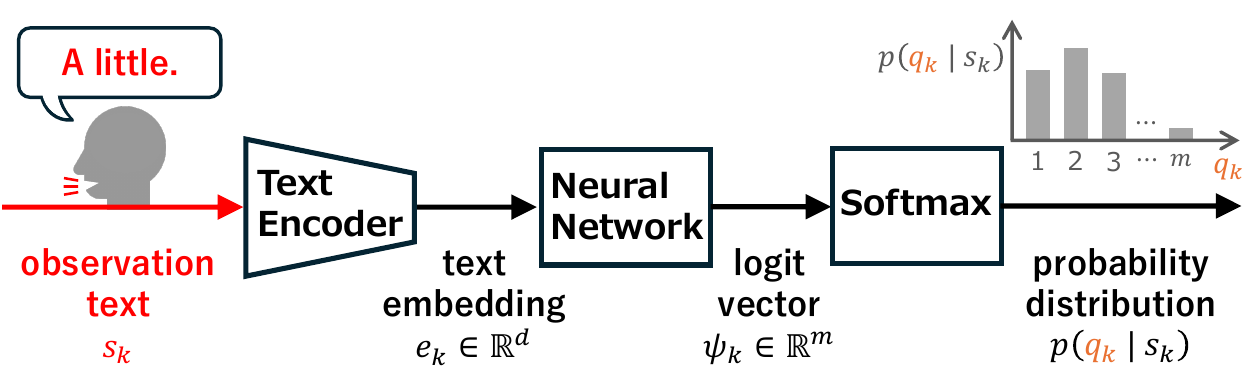}
		\caption{Block Diagram of the Model for Computing the Probability Distribution $p (q_k | s_k)$  \label{fig:Block Diagram of the Model for Computing the Probability Distribution}}
\end{figure}

\modifiedIV{\textbf{Computation of $p(q_k^j | s_k)$}:}
% Subsequently, we
% % explain how to compute
% \modifiedII{provide the computation of} $p(q_k^j | s_k)$.
% 本稿では，$p (q_k^j | s_k)$の計算のため，\figref{fig:Block Diagram of the Model for Computing the Probability Distribution}に示す量子ラベル分類モデルを提案する．
% For this purpose, 
We propose a quantized-label classification model, illustrated in \figref{fig:Block Diagram of the Model for Computing the Probability Distribution}.
% このモデル構造について説明する．
% We describe the structure of this model.
% まず，観測テキスト$s_k$をテキストエンコーダに入力し，テキスト埋め込み$e_k \in \mathbb{R}^d$を取得する．
First,
% the observation text $s_k$ is input to a text encoder to obtain a text embedding $e_k \in \mathbb{R}^d$.
\modifiedIV{a text encoder processes the observation text $s_k$ to produce an embedding $e_k \in \mathbb{R}^d$.}
% テキストエンコーダとは，入力されたテキストを意味や文脈を考慮した数値ベクトルに変換する機械学習モデルである．
A text encoder is a machine learning model that maps an input text into a vector representation capturing its semantics and context.
% 例として，テキスト埋め込みモデルや事前学習済みの言語モデルが挙げられ，得られるテキスト埋め込み$e_k$の次元数$d$は一般に数百から数千程度と高次元になる．
Examples include text embedding models or pretrained language models, and the dimensionality $d$ of the embedding $e_k$ is typically high, ranging from several hundred to several thousand.
% 次に，テキスト埋め込み$e_k$をニューラルネットワークに入力し，$m$次元の特徴量$\psi_k$を取得する．
Second, 
% the embedding $e_k$ is transformed by a neural network into an $m$-dimensional feature vector $\psi_k$.
\modifiedIV{a neural network maps the embedding $e_k$ to an $m$-dimensional feature vector $\psi_k$.}
% 最後に，softmax関数を適用することで，確率分布$p (q_k | s_k)$を得ることができる．
Finally,
% applying the softmax function to $\psi_k$ yields the probability distribution $p(q_k | s_k)$.
\modifiedIV{the softmax function applied to $\psi_k$ yields the probability distribution $p(q_k | s_k)$.}
% このモデルは，量子化ラベル$q$と観測テキスト$s$の組のデータセットを用いて，教師あり学習により訓練することができる．
This model can be trained using supervised learning with a dataset of text-label pairs $\{s^i, q^i\}_{i=1}^{N_{\mathrm{d}}}$.
% このデータセットは，\ref{subsec:Dataset}節に示すようなアンケートを実施することで収集することができる．
Such a dataset can be constructed by conducting questionnaires, as 
% described
\modifiedIV{detailed} in Subsection~\ref{subsec:Dataset}.
\modifiedIV{
\begin{rem}
    Computing the probability distribution $p(q_k | s_k)$, rather than directly predicting a single label $q_k$, has a benefit in handling out-of-domain observation texts $s_k$.
    For example, when an observation text $s_k$ contains features not present in the training data, the model outputs a low-confidence distribution instead of a forced high-confidence label, thereby preventing overconfident and incorrect likelihood updates.
    % 先行研究の話
    This idea is inspired by the approach of \cite{sitdhipol2025spatial}, which focuses on spatial-language expressions.
    In contrast, we provide a method applicable to general natural language observations.
\end{rem}
}

%% [段落3-1-5]p(q|x)の計算
% 最後に，$p (q_k^j | x_{k | k-1}^i)$の計算方法について述べる．
\modifiedIV{\textbf{Computation of $p(q_k^j | x_{k | k-1}^i)$}:}
% Finally, we
% % describe how to compute
% \modifiedII{provide the computation of} $p(q_k^j | x_{k | k-1}^i)$.
% $p (q_k | x_k)$について，次の定理が成り立つ．
The following
% theorem
\modifiedII{proposition} provides a formulation of $p(q_k | x_k)$:

\begin{prop} \label{prop:PriorDistributuon of q}
    % 状態が$x_k$である条件のもとで量子ラベルが$q_k$である確率$p (q_k | x_k)$は次式で記述できる．
    % The conditional probability $p(q_k | x_k)$ is obtained as
    \modifiedIV{It holds that}
    \begin{align}
            p (q_k | x_k) = \int_{\Lambda_{q_k}} p (y_{\mathrm{H}, k} | x_k) \; \mathrm{d} y_{\mathrm{H}, k}, \label{eq:Theo of PriorDistributuon of q}
    \end{align}
    % ただし，$\Lambda_{q_k}$は$i = q_k$の場合の$\Lambda_i$であり，量子ラベル$q_k$に対応する量子化区間である．
    where $\Lambda_{q_k}$ denotes the quantization interval corresponding to the label $q_k$, such that $\Lambda_{q_k} = \Lambda_i$ when $i = q_k$.
\end{prop}
\begin{pf} \label{Proof:PriorDistributuon of q}
    % 確率$p (q_k | x_k)$は，$p (q_k, y_{\mathrm{H}, k} | x_k)$を認知値$y_{\mathrm{H}, k}$について周辺化することで，次式で記述できる．
    % The probability $p(q_k | x_k)$ is obtained by marginalizing the joint distribution $p(q_k, y_{\mathrm{H}, k} | x_k)$ with respect to the cognitive value $y_{\mathrm{H}, k}$:
    \modifiedII{
    Note that the probability $p(q_k | x_k)$ is obtained by marginalizing the joint distribution $p(q_k, y_{\mathrm{H}, k} | x_k)$ with respect to the cognitive value $y_{\mathrm{H}, k}$.
    In addition, recall the definition of conditional probability and the fact that the quantized label $q_k$ depends only on $y_{\mathrm{H}, k}$ as follows from \eqref{subeq:Quantizer of LanguageGenerater 1}.
    Then, it follows that
    }
    \begin{align}
            p (q_k | x_k)
            % &= \int_{-\infty}^{\infty} p (q_k, y_{\mathrm{H}, k} | x_k) \; \mathrm{d} y_{\mathrm{H}, k} \nonumber \\
            % &= \int_{-\infty}^{\infty} p (q_k | y_{\mathrm{H}, k}, x_k)  \; p (y_{\mathrm{H}, k} | x_k) \; \mathrm{d} y_{\mathrm{H}, k} \nonumber \\
            % &= \int_{-\infty}^{\infty} p (q_k | y_{\mathrm{H}, k})  \; p (y_{\mathrm{H}, k} | x_k) \; \mathrm{d} y_{\mathrm{H}, k}.
            = \int_{-\infty}^{\infty} p (q_k | y_{\mathrm{H}, k})  \; p (y_{\mathrm{H}, k} | x_k) \; \mathrm{d} y_{\mathrm{H}, k}.
            \label{eq:Proof of PriorDistributuon of q 1}
    \end{align}
    % $1$行目から$2$行目への変形は，条件付き確率の定義によるものである．
    % The second equality follows from the definition of conditional probability.
    % $2$行目から$3$行目への変形が可能な理由は，\eqref{subeq:Quantizer of LanguageGenerater 1}式より量子ラベル$q_k$が認知値$y_{\mathrm{H}, k}$のみで決定し，$p (q_k | y_{\mathrm{H}, k}, x_k)=p (q_k | y_{\mathrm{H}, k})$と記述できるためである．
    % The third equality follows from the fact that  $p(q_k | y_{\mathrm{H}, k}, x_k) = p(q_k | y_{\mathrm{H}, k})$.  
    % This follows from \eqref{subeq:Quantizer of LanguageGenerater 1}, which states that the quantized label $q_k$ depends only on the cognitive value $y_{\mathrm{H}, k}$.

    % \eqref{eq:QuantizerFunc}式より，$p (q_k | y_{\mathrm{H}, k})$は次式で記述できる．
    From \eqref{eq:QuantizerFunc},
    % the conditional distribution
    $p(q_k | y_{\mathrm{H}, k})$ is given by
    \begin{equation} \label{eq:Proof of PriorDistributuon of q 2}
            p (q_k | y_{\mathrm{H}, k}) =
            \begin{cases}
                    1, & \mathrm{if} \; y_{\mathrm{H}, k} \in \Lambda_{q_k}, \\
                    0, & \mathrm{otherwise}.
            \end{cases}
    \end{equation}
    % \eqref{eq:Proof of PriorDistributuon of q 1}式に\eqref{eq:Proof of PriorDistributuon of q 2}式を代入することで，次式が導ける．
    By substituting \eqref{eq:Proof of PriorDistributuon of q 2} into \eqref{eq:Proof of PriorDistributuon of q 1},
    % we obtain
    % \modifiedII{it follows that}
    \modifiedIV{we have \eqref{eq:Theo of PriorDistributuon of q}.}
    % \begin{align}
    %         p (q_k | x_k) = \int_{\Lambda_{q_k}} p (y_{\mathrm{H}, k} | x_k) \; \mathrm{d} y_{\mathrm{H}, k}. \label{eq:Proof of PriorDistributuon of q 3}
    % \end{align}
    % これにより，\propref{prop:PriorDistributuon of q}が示された．
    % This completes the proof of \modifiedII{\propref{prop:PriorDistributuon of q}.}
\end{pf}
% \propref{prop:PriorDistributuon of q}より，$p (q_k^j | x_{k | k-1}^i)$は\eqref{eq:Theo of PriorDistributuon of q}式の積分を計算することで得ることができる．
% From \modifiedII{\propref{prop:PriorDistributuon of q},} the probability $p(q_k^j | x_{k | k-1}^i)$ is obtained by evaluating the integral in \eqref{eq:Theo of PriorDistributuon of q}.

%% [段落3-1-5]全体のアルゴリズムを示す
% 最後に，LAPFのアルゴリズムを\algref{alg:Language-Aided Particle Filter}に示す．
To conclude this subsection, the LAPF algorithm is presented in \textbf{Algorithm~\ref{alg:Language-Aided Particle Filter}}.
\begin{algorithm}[t]
\caption{Language-Aided Particle Filter}
\label{alg:Language-Aided Particle Filter}
\begin{algorithmic}[1]
\State \textbf{Initialization:}
\State Sample particles $x_{0 | 0}^i \sim \pi(x_0)$ for $i=1,\dots,N_p$
\State Set initial weights $\alpha_0^i = 1/N_p$
\For{$k = 1$ to $T$}
    \State $x_{k | k-1}^i \sim p(x_k | x_{k-1 | k-1}^i)$
    \State $p(s_k | x_{k | k-1}^i) \gets \sum_{j = 1}^m \; p (q_k^j | s_k)  \; p (q_k^j | x_{k | k-1}^i)$
    \State $\alpha_k^i \gets p(s_k | x_{k | k-1}^i)$
    \State $\alpha_k^i \gets \alpha_k^i / \sum_{j=1}^{N_p} \alpha_k^j$
    \State $\tilde{\pi}(x_k | s_{1 : k}) \gets \sum_{i=1}^{N_p} \alpha_k^i \, \delta(x_k - x_{k | k-1}^i)$
    \State Resample particles $x_{k | k}^i \sim \tilde{\pi}(x_k | s_{1 : k})$
\EndFor
\end{algorithmic}
\end{algorithm}
% ここで，\algref{alg:Language-Aided Particle Filter}に現れる$p(x_k | x_{k-1 | k-1})$は状態遷移モデルであり，\eqref{eq:PredictionStep of ParticleFilter1}式の状態方程式に対応している．
% Here,
\modifiedII{Note that} $p(x_k | x_{k-1 | k-1})$ in \textbf{Algorithm~\ref{alg:Language-Aided Particle Filter}} denotes the state-transition model, which corresponds to the state equation \eqref{eq:PredictionStep of ParticleFilter1}.

\subsection{The case of multiple human sensors}

%% [段落3-2-1]概要
% 本小節では，$r$個の人センサ$\mathcal{S}_{\mathrm{H}}$を使用する場合の尤度$p (s_k^{(1)}, \dots , s_k^{(r)} | x_{k | k-1}^i)$の計算方法について述べる．
In this subsection,
\modifiedIV{assuming multiple human sensing agents,}
we
% describe how to compute
\modifiedII{% provide
\modifiedIV{discuss} the computation of} the likelihood $p (s_k^{(1)}, \dots , s_k^{(\modifiedIV{N_{\mathrm{H}}})} | x_{k | k-1}^i)$ when using \modifiedIV{$N_{\mathrm{H}}$ agents.}
% human sensors $\mathcal{S}_{\mathrm{H}}$.

%% [段落3-2-2]複数の場合の尤度
% 同時確率$p (s_k^{(1)}, \dots , s_k^{(r)} | x_{k | k-1}^i)$は，条件付き確率分布の連鎖律により次式で記述できる．
% The joint probability $p(s_k^{(1)}, \dots , s_k^{(r)} | x_{k | k-1}^i)$ is obtained as
\modifiedIII{By applying the chain rule of conditional probability, the joint probability $p(s_k^{(1)}, \dots , s_k^{(\modifiedIV{N_{\mathrm{H}}})} | x_{k | k-1}^i)$ holds that}
% \begin{align}
%     p (s_k^{(1)},& \dots , s_k^{(r)} | x_{k | k-1}^i) \nonumber \\
% 		&= \prod_{j = 1}^r \; p (s_k^{(j)} | s_k^{(1)}, \dots , s_k^{(j-1)}, x_{k | k-1}^i), \label{eq:How to get Likelihood in Multi HumanSensor 2}
% \end{align}
\begin{equation} \label{eq:How to get Likelihood in Multi HumanSensor 2}
		\begin{array}{ll}
				p (s_k^{(1)},& \dots , s_k^{(\modifiedIV{N_{\mathrm{H}}})} | x_{k | k-1}^i) \\
				&= \prod_{j = 1}^{\modifiedIV{N_{\mathrm{H}}}} \; p (s_k^{(j)} | s_k^{(1)}, \dots , s_k^{(j-1)}, x_{k | k-1}^i),
		\end{array}
\end{equation}
% by applying the chain rule of conditional probability,
% ただし，$p (s_k^{(1)} | s_k^{(1)}, s_k^{(0)}, x_{k | k-1}^i) = p (s_k^{(1)} |  x_{k | k-1}^i)$である．
% Note that the first term satisfies $p(s_k^{(1)} | s_k^{(1)}, s_k^{(0)}, x_{k | k-1}^i) = p(s_k^{(1)} | x_{k | k-1}^i)$.
\modifiedIII{where $p(s_k^{(1)} | s_k^{(1)}, s_k^{(0)}, x_{k | k-1}^i) = p(s_k^{(1)} | x_{k | k-1}^i)$.}
% ここで，各観測テキスト$s_k^{(i)}$に対して, 次の仮定をおく．
% Here, we make the following assumption regarding the observation texts:
% \begin{assum} \label{asm:RecognizationNoise}
%     % 観測テキスト$[s_k^{(1)} \; \cdots \; s_k^{(r)}]^\top$において，各観測テキスト$s_k^{(i)}$は独立であると仮定する．
% 		In the observation vector $[s_k^{(1)} \; \cdots \; s_k^{(\modifiedIV{N_{\mathrm{H}}})}]^\top$, each observation text $s_k^{(i)}$ is assumed to be independent.
% \end{assum}
\modifiedIV{Assuming that the observation texts $s_k^{(j)}$ are independent of each other, \eqref{eq:How to get Likelihood in Multi HumanSensor 2} simplifies to}
% \asmref{asm:RecognizationNoise}より，\eqref{eq:How to get Likelihood in Multi HumanSensor 2}式は次式に書き換えられる．
% Equation~\eqref{eq:How to get Likelihood in Multi HumanSensor 2} can be simplified to the following form by \asmref{asm:RecognizationNoise}:
\begin{align}
    p (s_k^{(1)}, \dots , s_k^{(\modifiedIV{N_{\mathrm{H}}})} | x_{k | k-1}^i) = \prod_{j = 1}^{\modifiedIV{N_{\mathrm{H}}}} \; p (s_k^{(j)} | x_{k | k-1}^i). \label{eq:How to get Likelihood in Multi HumanSensor 3}
\end{align}
% ここで，$p (s_k^{(j)} | x_{k | k-1}^i)$は，\ref{subsec:How to get Likelihood}節で考えていた尤度$p (s_k | x_{k | k-1}^i)$であり，\eqref{eq:How to get Likelihood3}式で計算できる．
% Here, each term
\modifiedIV{Note that each} $p (s_k^{(j)} | x_{k | k-1}^i)$ corresponds to the single-sensor likelihood $p (s_k | x_{k | k-1}^i)$
% considered
\modifiedIV{addressed} in Subsection~\ref{subsec:How to get Likelihood} and
% can be computed using
\modifiedIV{is given by} \eqref{eq:How to get Likelihood3}.
% そのため，複数の人センサ$\mathcal{S}_{\mathrm{H}}$を使用する場合の尤度は，各観測テキスト$s_k^{(j)}$に対して尤度$p (s_k^{(j)} | x_{k | k-1}^i)$を計算した後で，それらの総積をとることで計算することができる．
Therefore, in the case of multiple human sensors $\mathcal{S}_{\mathrm{H}}$, the overall likelihood is obtained by computing the likelihood for each observation text $s_k^{(j)}$ and taking the product of these terms.

\section{Numerical Experiment} \label{sec:Numerical Experiment}

%% [段落4-1]概要
\modifiedIII{
% 本節では，数値実験でLAPFの有効性について述べる．
In this section, we demonstrate the effectiveness of the proposed LAPF through numerical experiments.
% そのために、irrigation canal の水位の状態推定問題に取り組む
To this end, we address a water level state estimation problem in an irrigation canal, 
% as
\modifiedIV{which is} illustrated in \figref{fig:River_Model_Eng}.
\begin{figure}
    \centering
    \includegraphics[width=0.6\linewidth]{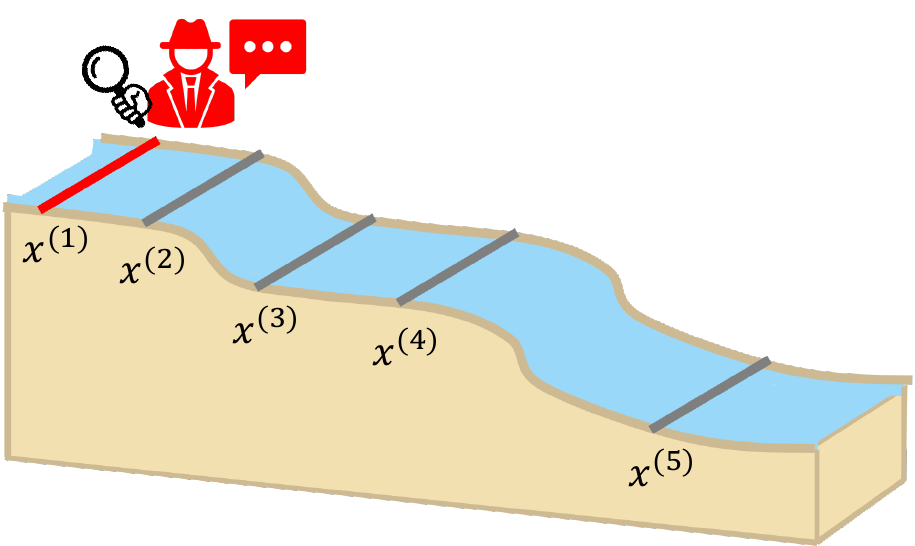}
    \caption{Schematic of the Canal Setting with Observation Points}
    \label{fig:River_Model_Eng}
\end{figure}
% 特に農業で田畑に水を適切に供給することが求められる canal では、その水位を正確に予測して制御する必要がある
In such canals, accurate prediction and control of the water level are crucial to ensure appropriate water allocation to agricultural fields.
% しかしながら、多くの場合には物理センサがない（Maestreさんの論文）何らかの代替手段で状態推定が必要である
However, in many practical situations, physical sensors are only sparsely deployed, as pointed out by \cite{sadowska2015human}.
Therefore, rather than relying
\modifiedIV{solely} on physical sensors,
% there is a need for alternative means of performing state estimation.
\modifiedIV{the state estimation that utilizes additional sources of information is needed.}
% 本研究では農家など人による観測のもと、Canalの物理モデルと組み合わせた状態推定問題に取り組む
In this paper, we address the problem of estimating the canal water level by combining a physical model of the canal with human observations from farmers,
\modifiedIV{and present the LAPF approach.}
% ベンチマーク相手として、別のAI-aidedな状態推定法である DNN-... と比較して、本提案法の有用性をあきらかにする。
To benchmark the LAPF, we additionally perform experiments with an External DNN-Aided Particle Filter (EDAPF).
The EDAPF was devised as a reference method, inspired by the concept of external deep-neural-network(DNN) architectures introduced in \cite{shlezinger2024ai}.
}

% % 本節では，LAPFの推定精度を調べた数値実験について述べる．
% In this section, we present numerical experiments conducted to evaluate the estimation accuracy of the proposed LAPF by comparing it with an External DNN-Aided Particle Filter (EDAPF).
% % 具体的には，\cite{shlezinger2024ai}で紹介された External DNN Architectures を参考に考案した External DNN Aided ParticleFilter と推定精度の比較を行なった．
% The EDAPF was devised as a reference method, inspired by the concept of external deep-neural-network(DNN) architectures introduced in \cite{shlezinger2024ai}.
% % , since no existing approach could be directly applicable to our setting.
% % 状況設定としては，\figref{fig:Schematic of the River Setting}に示す河川を想定し，その水位の推定を行なった．
% % The experimental scenario assumes a river environment, as illustrated in \figref{fig:Schematic of the River Setting}, where the task is to estimate the water level.
% % \begin{figure}[t]
% % 		\centering
% % 		\includegraphics[width=0.55\linewidth]{figure/River_Image_Eng.pdf}
% % 		\caption{Schematic of the River Setting \label{fig:Schematic of the River Setting}}
% % \end{figure}
% As the experimental scenario, we considered river water level estimation.

\subsection{Experiment Setting} \label{subsec:Experiment Setting}

%% [段落4-1-1]概要
% 本小節では，本実験における真のシステムの状態空間モデルと数値設定について述べる．
In this subsection, we describe the ground-truth state-space model and numerical settings used in our experiments.

%% [段落4-1-2]対象システム
% 対象システム$\mathcal{P}$として，次式の河川の水量伝播モデルを考えた．
We consider the following river flow propagation model as the \modifiedI{plant} system $\mathcal{P}$:
\begin{align}
    \mathcal{P}: \; x_k = \mathrm{proj}_{[0,5]} (A x_{k-1} + w_{k}), \quad \forall \; k \ge 1, \label{eq:StateEquation in Experiment}
\end{align}
% ここで，状態$x_k = [x_k^{(1)} \; \cdots \; x_k^{(5)}]^\top$は$5$つの地点の水位，関数$\mathrm{proj}_{[a, b]}:\mathbb{R}^n \to \mathbb{R}^n$は，ベクトルの各成分に対して区間$[a, b]$への直交射影を行なう関数である．
% Here,
\modifiedIV{where} the state vector $x_k = [x_k^{(1)} \; \cdots \; x_k^{(5)}]^\top$ represents the water levels at five locations, and the function $\mathrm{proj}_{[a,b]}:\mathbb{R}^n \to \mathbb{R}^n$ denotes the component-wise projection of a vector onto the interval $[a,b]$.
% したがって，\eqref{eq:StateEquation in Experiment}式は各地点の最高水位が$5$である河川の水位の時間発展を記述したものである．
Thus, 
% Equation~\eqref{eq:StateEquation in Experiment}
\modifiedIV{the model~\eqref{eq:StateEquation in Experiment}} describes the temporal evolution of river water levels, where the maximum water level at each location is bounded by~$5$.
% システム行列$A$の数値を次式に示す．
The system matrix $A$ is given by
\begin{align}
    A =
    \begin{bmatrix}
        0.4 & 0 & 0 & 0 & 0 \\
        0.6 & 0.3 & 0 & 0 & 0 \\
        0 & 0.7 & 0.5 &0 & 0 \\
        0 & 0 & 0.5 & 0.4 & 0 \\
        0 & 0 & 0 & 0.6 & 0.5
    \end{bmatrix}. \label{eq:System Matrix}
\end{align}
% システム雑音$w_k$は，$u_k = [1\;0\;0\;0\;0]^\top$を平均，$Q = 0.1 I_5$を共分散行列とした正規分布$\mathcal{N}(u_k, Q)$に従うとした．
The process noise $w_k$ is modeled as a Gaussian distribution $\mathcal{N}(u_k, Q)$ with mean $u_k = [1\;0\;0\;0\;0]^\top$ and covariance matrix $Q = \mathrm{diag}(1.0,0.1,0.1,0.1,0.1)$.
% また，各地点の初期水位$x_0^{(i)}$は$2.5$とした．
The initial water level at each location $x_0^{(i)}$ is set to $2.5$.

%% [段落4-1-3]人センサ
% 人センサ$\mathcal{S}_{\mathrm{H}}$として，次式を考えた．
\modifiedIII{We consider the following model as the ground-truth human sensing process in
% our simulation,
\modifiedIV{the experiments,} while the estimator relies on the observation model given in \eqref{eq:Recognizer of HumanSensor 1}
% ,
\modifiedIV{and} \eqref{eq:LanguageGenerater of HumanSensor 1}:}
% We consider the following model for the human sensor $\mathcal{S}_{\mathrm{H}}$:
\begin{subequations} \label{eq:HumanSensor in Experiment}
    \begin{empheq}[left={\mathcal{S}_{\mathrm{H}}: \empheqlbrace}]{alignat=2}
        y_{\mathrm{H}, k} &= \mathrm{proj}_{[0,5]} (C_{\mathrm{H}} x_k + v_{\mathrm{H}, k}),& \; \; \, & \forall \; k \ge 1, \label{subeq:Recognizer in Experiment} \\
        s_k &\sim \mathrm{Y2Prob} (y_{\mathrm{H}, k}),& &\forall \; k \ge 1, \label{subeq:LanguageGenerater in Experiment}
    \end{empheq}
\end{subequations}
% \eqref{subeq:Recognizer in Experiment}式において直交射影関数$\mathrm{proj}_{[0,5]}$を使っている理由は，河川の様子を観測した人間がその水位の下限・上限を越えて認知することはないと考えられるためである．
% Here, 
\modifiedIV{where} the projection operator $\mathrm{proj}_{[0,5]}$ in \eqref{subeq:Recognizer in Experiment} ensures that the perceived water level remains within the range $[0,5]$, reflecting the fact that
% a human observer
\modifiedI{an agent}
cannot perceive values outside this range.
% 認知行列$C_{\mathrm{H}}$は$[1\;0\;0\;0\;0]$とした．
% つまり，観測者は，状態$x_k^{(1)}$のみ観測可能とした．
The cognitive matrix 
\modifiedIV{$C_{\mathrm{H}}$}
is set to
% $C_{\mathrm{H}} = [1\;0\;0\;0\;0]$, 
\modifiedIV{$[1\;0\;0\;0\;0]$,} meaning that the
% observer
\modifiedI{agent}
perceives only the water level at the first location $x_k^{(1)}$.
% 認知雑音$v_{\mathrm{H}, k}$は，正規分布$\mathcal{N}(0, 0.5)$に従うとした．
The cognitive noise $v_{\mathrm{H}, k}$ is modeled as Gaussian with distribution $\mathcal{N}(0,1.0)$.
% \eqref{subeq:LanguageGenerater in Experiment}式における関数$\mathrm{Y2Prob}: \mathbb{R} \to \mathbb{R}^{N_{\mathrm{d}}}$は，認知値$y_{\mathrm{H}, k}$に対応する観測テキスト$s_k$の確率分布を出力する関数である．
In \eqref{subeq:LanguageGenerater in Experiment}, the function $\mathrm{Y2Prob}: \mathbb{R} \to \mathbb{R}^{N_{\mathrm{d}}}$ generates a probability distribution over possible observation texts $s_k$ corresponding to the cognitive value $y_{\mathrm{H}, k}$.
% \eqref{subeq:LanguageGenerater in Experiment}式の具体的な実装は，収集した認知値と観測テキストの$N_{\mathrm{d}}$組のデータセット$\{y_{\mathrm{H}}^i, s^i\}_{i=1}^{N_{\mathrm{d}}}$を使い，入力の認知値$y_{\mathrm{H}, k}$に対応する観測テキストデータ$s^i$の中からランダムに一つ選択することにより行なった．
In practice, we implement this function with a dataset of $N_{\mathrm{d}}$ pairs $\{y_{\mathrm{H}}^i, s^i\}_{i=1}^{N_{\mathrm{d}}}$ \modifiedIV{: given}
% .
% Given
a cognitive value $y_{\mathrm{H}, k}$, the corresponding observation text $s_k$ is obtained by randomly sampling one of the texts $s^i$ linked to
% that value.
\modifiedIV{the value of $y_{\mathrm{H}, k}$}
% このデータセットの詳細については，次の小節で述べる．
The details of this dataset are provided in Subsection~\ref{subsec:Dataset}.

\subsection{Dataset for Numerical Experiments} \label{subsec:Dataset}

%% [段落4-2-1]作成したデータセットについて
% Yahoo! クラウドソーシングにおいて，水位の割合を与え，その水位の河川を見たときにSNSにどのような投稿をするかについてアンケートを実施した．
We conducted a questionnaire survey on Yahoo! Crowdsourcing, an online platform in Japan, asking participants to imagine posting on SNS when observing a river at a given water level ratio.
% この調査は日本で行ない，テキストデータも日本語．
% The survey was conducted in Japanese, and thus the collected observation texts are originally written in Japanese.
% 得られた河川の水位の割合と観測テキストのデータの例を\tabref{tab:TextData}に示す．
Examples of the resulting pairs of water level ratios and observation texts
% , translated into English,
are shown in Table~\ref{tab:TextData}.
\begin{table}[t]
    \centering
    \caption{
    Dataset Example (English Translations): Water Levels and Texts
    }
    \begin{tabular}{c|c}
    \hline
    Water level Ratio & Observation Text \\
    \hline
    4$\%$ & There’s barely any water out here. \\
		40$\%$ & The river’s flowing really gently today. \\
    80$\%$ & The water’s pretty high… hope it’s okay. \\
    98$\%$ & Almost flooding… this is scary! \\
    \hline
    \end{tabular}
    \label{tab:TextData}
\end{table}
% 水位の割合と観測テキストのデータを，水位の割合が$0\%$から$100\%$まで$2\%$区切りで各水位$50$人分取得し，最終的に計$2378$組のデータセットを作成した．
We collected observation texts from $50$ participants for each water level ratio between $0\%$ and $100\%$ in increments of $2\%$, resulting in a dataset of $2,454$ pairs.
% そして，このデータセットのうち，$1882$組を訓練データ，$205$組を検証データ，$289$組をテストデータとした．
Among these, $1,882$ pairs were used for training, $205$ for validation, and $289$ for testing.
% 訓練データと検証データは，\figref{fig:Block Diagram of the Model for Computing the Probability Distribution}の量子ラベル分類モデルの訓練に使用し，テストデータを\eqref{subeq:LanguageGenerater in Experiment}式の言語化処理の実装に使用した．
The training and validation sets were used to train the quantized-label classification model in LAPF and an external DNN in EDAPF.
The test set was used to implement the language generation process of the true system described in \eqref{subeq:LanguageGenerater in Experiment}.

\subsection{Training of the Quantized-Label Classification Model} \label{subsec:Training of the Quantized-Label Classification Model}

%% [段落4-3-1]概要
% 本小節では，LAPFで使う量子ラベル分類モデルの訓練について話すよ
In this subsection, we describe the training of the quantized-label classification model used in LAPF. 
% 本実験では，得られた観測テキスト$s_k$から，認知値$y_{\mathrm{H}, k}$が，区間$\Lambda=[0,5]$を$5$等分した量子化区間$\{\Lambda_i\}_{i=1}^5$のどの区間に含まれているかを予測する量子ラベル分類モデルを構築・訓練した．
In the experiments, we constructed and trained a quantized-label classification model that predicts, from the observation text $s_k$, which of five quantization intervals the cognitive value $y_{\mathrm{H}, k}$ belongs to.  
Specifically, the range $\Lambda=[0,5]$ was divided into five equal partitions $\{\Lambda_i\}_{i=1}^5$, and the model assigns $y_{\mathrm{H}, k}$ to the corresponding $\Lambda_i$.

%% [段落4-3-2]モデル構造
% 構築した量子ラベル分類モデルについて述べる．
We describe the structure of the quantized-label classification model,
\modifiedIII{illustrated in \figref{fig:Block Diagram of the Model for Computing the Probability Distribution}}.
% テキストエンコーダとして，日本語のテキスト埋め込みモデル``sentence-bert-base-ja-mean-tokens-v2''\textsuperscript{\cite{web:sentence-bert}}を使用した．
% このモデルでは，入力されたテキスト$s_k$を$768$次元のテキスト埋め込み$e_k$に変換できる．
We used the Japanese embedding model ``sentence-bert-base-ja-mean-tokens-v2''\footnote{https://huggingface.co/sonoisa/sentence-bert-base-ja-mean-tokens-v2} 
as the text encoder, which maps an input text $s_k$ to a text embedding $e_k \in \mathbb{R}^{768}$. 
% テキスト埋め込み$e_k$から特徴量$\psi_k \in \mathbb{R}^5$を出力するニューラルネットワークとして，次式を構築した．
% \begin{subequations} \label{eq:Classification Model}
%     \begin{empheq}[]{alignat=2}
%         \theta_{1, k} &= \mathrm{ReLU}(W_1 e_k + b_1) \\
% 				\theta_{2, k} &= \mathrm{ReLU}(W_2 \theta_{1, k} + b_2) \\
% 				\psi_k &= W_3 \theta_{2, k} + b_3
%     \end{empheq}
% \end{subequations}
% ここで，$\mathrm{ReLU}: \mathbb{R}^d \to \mathbb{R}^d$は活性化関数ReLUである．
% 各パラメータの次元数については，$W_1 \in \mathbb{R}^{128 \times 768}$，$b_1 \in \mathbb{R}^{128}$，$W_2 \in \mathbb{R}^{64 \times 128}$，$b_2 \in \mathbb{R}^{64}$，$W_3 \in \mathbb{R}^{5 \times 64}$，$b_3 \in \mathbb{R}^{5}$とした．
This embedding is then processed by a neural network consisting of two hidden layers
\modifiedIII{
of 128 and 64 units, each followed by a ReLU activation.
The final linear layer maps the 64-dimensional representation to a 5-dimensional feature vector $\psi_k \in \mathbb{R}^5$, which is converted into a probability distribution over the five quantized labels via a softmax function.
}
% The first layer applies a ReLU activation after a linear transformation with weight matrix $W_1 \in \mathbb{R}^{128 \times 768}$ and bias $b_1 \in \mathbb{R}^{128}$. 
% The second layer applies another ReLU activation after a linear transformation with $W_2 \in \mathbb{R}^{64 \times 128}$ and $b_2 \in \mathbb{R}^{64}$. 
% Finally, a linear transformation with $W_3 \in \mathbb{R}^{5 \times 64}$ and $b_3 \in \mathbb{R}^5$ outputs the feature vector $\psi_k \in \mathbb{R}^5$. 
% Applying the softmax function to $\psi_k$ yields the probability distribution over the five quantized labels.

%% [段落4-3-3]訓練の話
% 構築したモデルの訓練について述べる．
% We describe the training of this model. 
% % 訓練には\ref{subsec:Dataset}節で述べた訓練データを用いた．
% The training data described in Subsection~\ref{subsec:Dataset} were used. 
% 訓練対象のパラメータは，ニューラルネットワークの$W_1$，$b_1$，$W_2$，$b_2$，$W_3$，$b_3$のみとした．
The trainable parameters were limited to those of the neural network.
% , namely $W_1$, $b_1$, $W_2$, $b_2$, $W_3$, and $b_3$. 
% 損失関数にはクロスエントロピー損失を用いた．
% 最適化には Adam を使い，初期学習率を$1.0 \times 10^{-5}$に設定した．
The model was trained using the cross-entropy loss function and optimized with Adam, with an initial learning rate of $1.0 \times 10^{-5}$. 
% バッチサイズは$16$，エポック数は$100$とた．
The batch size was set to $16$, and training was performed for $100$ epochs.

\subsection{Result} \label{subsec:Result}

\modifiedIII{
%% [段落4-5-0]そのほかの設定
% LAPFとEDAPFのどちらの状態推定でも，初期時刻$k=0$における近似事後分布$\tilde{\pi} (x_0)$は，平均が初期状態の真値$x_0$，共分散行列が単位行列$I_5$の正規分布$\mathcal{N}(x_0, I_5)$とした．
For both the LAPF and the EDAPF,
% the approximate posterior distribution at the initial time $k=0$, denoted by $\tilde{\pi}(x_0)$, was initialized as a Gaussian distribution $\mathcal{N}(x_0, I_5)$ with mean equal to the true initial state $x_0$ and covariance matrix $I_5$.
\modifiedVII{
we let $\pi(x_0) = \mathcal{N}(0, I_5)$.
}%
% また，総時間ステップ数は$100$，パーティクルの数$N_{\mathrm{p}}$は$1000$とした．
The total number of time steps was set to $100$, and the number of particles $N_{\mathrm{p}}$ was set to $1,000$.
% 以上の設定における状態推定を両者$1000$回行なった．
Under these settings, state estimation was performed $1,000$ trials for each method.
The implementation details of the EDAPF are provided in Appendix~\ref{subsec:Comparative Method: EDAPF}.
}

%% [段落4-5-1]結果1
% 推定値$\hat{x}k$は経験事後分布$\tilde{\pi}(x_k | s_{1:k})$の平均と定義
% For both methods,
\modifiedIII{In this experiment,}
we defined the state estimate $\hat{x}_k$ as the mean of the empirical distribution $\tilde{\pi}(x_k | s_{1:k})$.
% 全試行・全時刻にわたる状態 $x_k$ の推定値と真値との間のMSEを計算した．
We computed
% the MSE
\modifiedIV{the mean squared error (MSE)} between $\hat{x}_k$ and the true states $x_k$
% across all trials and time steps.
\modifiedIV{to evaluate the estimation accuracy.}
\begin{figure}[t]
		\centering
		\includegraphics[width=0.78\linewidth]{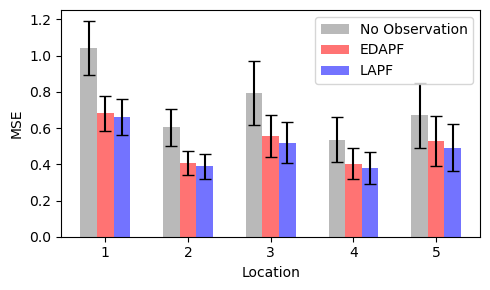}
		\caption{Comparison of State-wise MSE: Observation-Free vs. EDAPF vs. LAPF \label{fig:Comparison of MSE}}
\end{figure}
\modifiedIII{
As shown in \figref{fig:Comparison of MSE}, we first evaluated a baseline model that performs state estimation without any observations,
% using only the time update of the physical model.
\modifiedIV{meaning the state estimation that relies solely on the plant model \eqref{eq:StateEquation in Experiment}.}
This observation-free setting resulted in an average MSE of $0.73 \pm 0.13$ (mean ± standard deviation across trials).
Both the LAPF and EDAPF achieved lower MSE than this baseline, indicating that human observations can be effectively exploited to improve estimation accuracy.
Furthermore, the LAPF consistently achieved lower MSE than the EDAPF across all locations.}
% % 全状態において，LAPFの方がEDAPFよりMSEが小さい
% The results, shown in \figref{fig:Comparison of MSE}, indicate that the LAPF consistently achieved lower MSE than the EDAPF across all locations.
% 試行ごとのMSEの平均とばらつき
Over all trials, the average MSE was $0.49 \pm 0.08$ for the LAPF and $0.52 \pm 0.08$ for the EDAPF.

%% [段落4-5-2]結果2
\modifiedIII{
% In an additional experiment, we evaluated the robustness of the method under out-of-domain observation texts.
\modifiedIV{We performed an additional experiment to evaluate the robustness of the LAPF under ``out-of-domain'' observation texts.}
To this end, we injected dialectal expressions,
\modifiedIV{which is not included in the training data,} as out-of-domain observations only when $y_{\mathrm{H},k} < 0.2$.
\begin{figure}[t]
		\centering
		\includegraphics[width=0.78\linewidth]{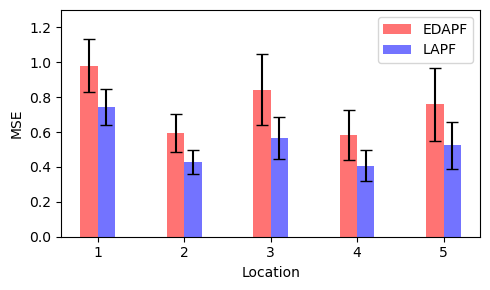}
		\caption{Comparison of State-wise MSE: EDAPF vs. LAPF (Out-of-domain) \label{fig:Comparison of MSE (OOD)}}
\end{figure}
As shown in \figref{fig:Comparison of MSE (OOD)}, the LAPF retained lower MSE than the EDAPF even under out-of-domain observations, indicating higher robustness to domain shifts in the observation texts.
Over all trials, the average MSE was $0.53 \pm 0.08$ for the LAPF and $0.75 \pm 0.15$ for the EDAPF.
}

\modifiedIII{
This robustness can be attributed to the different ways in which the two methods incorporate observation texts $s_k$.
While the LAPF uses a probability distribution $p(q_k | s_k)$ inferred from the observation text, the EDAPF uses only a single predicted value of $y_{\mathrm{H},k}$.
For example, 
% when an OOD dialectal expression was injected at $y_{\mathrm{H},k} < 0.2$,
\modifiedIV{consider that 
an observation text with dialectal expression is injected.}
\modifiedIV{Then,}
the LAPF produced a distribution such as $\begin{bmatrix}0.34 & 0.19 & 0.11 & 0.15 & 0.21\end{bmatrix}$, 
% whereas
\modifiedIV{while} the EDAPF yielded a single predicted value of $2.45$.
This suggests that representing observations as a distribution rather than a single point estimate contributes to the higher robustness of the LAPF under domain shifts.
}

\section{Conclusion} \label{sec:Conclusion}

%% [段落5-1]結論
% 本稿では，一般的な自然言語表現を活用可能にしたパーティクルフィルタとして，LAPF を提案した．
This paper 
% has presented
\modifiedI{presented} the LAPF, a particle filter framework that enables the use of general 
% natural language expressions
\modifiedI{natural language observations by human agents} in state estimation.
% LAPFとEDAPFの状態推定精度の比較実験を行なった結果，LAPFの方がEDAPFよりも推定精度が高くなった．
We conducted a comparative experiment between the LAPF and the EDAPF, and the results demonstrated that the LAPF achieved higher estimation accuracy.
\modifiedIII{Furthermore, in the additional experiment under out-of-domain observation texts, the LAPF exhibited smaller degradation in performance than the EDAPF, indicating its higher robustness to domain shifts in observation texts.}

\bibliography{ifacconf}             % bib file to produce the bibliography
                                                     % with bibtex (preferred)
                                                   
%\begin{thebibliography}{xx}  % you can also add the bibliography by hand

%\bibitem[Able(1956)]{Abl:56}
%B.C. Able.
%\newblock Nucleic acid content of microscope.
%\newblock \emph{Nature}, 135:\penalty0 7--9, 1956.

%\bibitem[Able et~al.(1954)Able, Tagg, and Rush]{AbTaRu:54}
%B.C. Able, R.A. Tagg, and M.~Rush.
%\newblock Enzyme-catalyzed cellular transanimations.
%\newblock In A.F. Round, editor, \emph{Advances in Enzymology}, volume~2, pages
%  125--247. Academic Press, New York, 3rd edition, 1954.

%\bibitem[Keohane(1958)]{Keo:58}
%\newblock \emph{Power and Interdependence: World Politics in Transitions}.
%\newblock Little, Brown \& Co., Boston, 1958.

%\bibitem[Powers(1985)]{Pow:85}
%\newblock Is there a way out?
%\newblock \emph{Harpers}, pages 35--47, June 1985.

%\bibitem[Soukhanov(1992)]{Heritage:92}
%\newblock \emph{{The American Heritage. Dictionary of the American Language}}.
%\newblock Houghton Mifflin Company, 1992.

%\end{thebibliography}

\appendix
\section{Comparative Method: EDAPF} \label{subsec:Comparative Method: EDAPF}
%% [段落4-4-1]EDAPFの概要
% 本小説では，LAPFの比較対象として，EDAPFを紹介する．
In this section, we introduce the EDAPF as a comparative method to the LAPF.
% EDAPFでは，external DNNを使って，観測テキスト$s_k$から直接認知値$y_{\mathrm{H},k}$を予測し，その予測値$\tilde{y}_{\mathrm{H},k}$を擬似的な観測値として用いてパーティクルフィルタを実行する方法である．
The EDAPF is a method that uses an external DNN to predict the cognitive value $y_{\mathrm{H},k}$ from the observation text $s_k$ directly, and treats the predicted value $\tilde{y}_{\mathrm{H},k}$ as a pseudo-observation for running the particle filter.
% 具体的には，予測値$\tilde{y}_{\mathrm{H},k}$を次の観測方程式の観測値とみなし，パーティクルフィルタを行なう．
Specifically, the predicted value $\tilde{y}_{\mathrm{H},k}$ is regarded as the observation in the following observation equation, and the particle filter is applied:
\begin{align}
    \tilde{\mathcal{S}}_{\mathrm{H}}: \; \tilde{y}_{\mathrm{H},k} = h_{\mathrm{H}}(x_k, v_{\mathrm{H},k}) + \tilde{v}_k, \quad \forall \; k \ge 1. \label{eq:PseudoObservationEquation 1}
\end{align}
% ここで，$\tilde{v}_k$は，external DNNによる予測誤差を保証するための追加の雑音である．
Here, $\tilde{v}_k$ denotes an additional noise term introduced to account for the prediction error of the external DNN.
% 本実験では，$\tilde{v}_k$は正規分布$\mathcal{N}(0, \tilde{R})$に従うとした．ただし，$\tilde{R}$はexternal DNNの平均二乗誤差
In this experiment, we modeled $\tilde{v}_k$ as a Gaussian distribution $\mathcal{N}(0, \tilde{R})$, where $\tilde{R}$ was set to the MSE of the external DNN evaluated on the validation data.

%% [段落4-4-2]external DNNのモデル構造
% 構築したexternal DNNについて述べる．
We describe the structure of the external DNN.
% まず，テキストをテキスト埋め込みに変換
First, the observation text $s_k$ is mapped to a text embedding $e_k \in \mathbb{R}^{768}$ using the Japanese embedding model ``sentence-bert-base-ja-mean-tokens-v2.''
% テキスト埋め込みをNNに入力
The embedding is then passed through a neural network with two hidden layers of 128 and 64 units, each followed by a ReLU activation.
% 最後シグモイド
% Finally, a linear transformation with $W_3^{\prime} \in \mathbb{R}^{1 \times 64}$ and $b_3^{\prime} \in \mathbb{R}$ outputs a scalar feature $\psi_k^{\prime} \in \mathbb{R}$, which is passed through a sigmoid function and scaled by the maximum water level of~$5$ to produce the prediction $\tilde{y}_{\mathrm{H},k}$.
A final linear layer outputs a scalar feature $\psi_k' \in \mathbb{R}$, which is passed through a sigmoid function and scaled by the maximum water level of $5$ to produce the prediction $\tilde{y}_{\mathrm{H},k}$

%% [段落4-4-3]external DNNの訓練
% 構築したモデルの訓練について述べる．
% We describe the training of this model. 
% % ほぼ量子ラベル分類モデルの時と同じ
% The external DNN was trained under almost the same conditions as the quantized-label classification model described in Subsection~\ref{subsec:Training of the Quantized-Label Classification Model}.
% 具体的に同じというのは
% Specifically,
% the trainable parameters were $W_1'$, $b_1'$, $W_2'$, $b_2'$, $W_3'$, and $b_3'$,
The trainable parameters were limited to those of the neural network,
and the optimizer was Adam with an initial learning rate of $1.0 \times 10^{-5}$.
The batch size was $16$, and training was performed for $100$ epochs.
% 唯一違うところは
The only difference is that the MSE loss function was used instead of the cross-entropy loss.
% 検証結果
The trained model achieved an MSE of $0.040$ on the validation dataset.

\end{document}